\documentclass{article}
\usepackage[preprint]{spconfa4}
\usepackage{amsmath,graphicx}
\usepackage{./echodefs}
\usepackage[lofdepth,lotdepth]{subfig}
\usepackage{hyperref}
\usepackage[font={footnotesize}]{caption}

\newcommand{\toeplitz}{\mathsf{Toeplitz}}
\renewcommand{\diag}{\mathsf{diag}}

\copyrightnotice{978-1-5386-8151-0/18/\$31.00~\copyright2018 IEEE}

\title{COCKTAILS, BUT NO PARTY: MULTIPATH-ENABLED PRIVATE AUDIO}
\name{Yu-Jeh Liu\textsuperscript{\dag}, Jonah Casebeer\textsuperscript{*}, and Ivan Dokmani\'{c}\textsuperscript{\dag}\thanks{The authors gratefully acknowledge  support from a Google Faculty Research Award ``Echonomy in auditory scene analysis''.}}
\address{\textsuperscript{\dag}Department of Electrical and Computer Engineering and \textsuperscript{*}Department of Computer Science \\ University of Illinois at Urbana-Champaign}
\begin{document}
\ninept
\maketitle
\begin{abstract}
We describe a private audio messaging system that uses echoes to unscramble messages at a few predetermined locations in a room. The system works by splitting the audio  into short chunks and emitting them from different loudspeakers. The chunks are filtered so that as they echo around the room, they sum to noise everywhere except at a few chosen focusing spots where they exactly reproduce the intended messages. Unlike in the case of standard personal audio zones, the proposed method renders sound outside the focusing spots unintelligible. Our method essentially depends on echoes: the room acts as a mixing system such that at given points we get the desired output. Finally, we only require a modest number of loudspeakers and only a few impulse response measurements at points where the messages should be delivered. We demonstrate the effectiveness of the proposed method via objective quantitative metrics as well as informal listening experiments in a real room. 
\end{abstract}

\section{Introduction}
\label{sec:intro}

Consider the following communication problem: using a set of loudspeakers in a room we want to transmit an audio message to Gwenda at point A and a different audio message to Waldemar at point B. Gwenda should not be able to understand Waldemar's message, nor should Waldemar understand Gwenda's. Crucially, no one else in the room should understand any of the messages.

This problem is related to personal audio zones and soundfield reproduction \cite{Betlehem_2015, Berkhout_1993, Ward_2001} where the requirement that the message should only be intelligible by its target recipient is usually not emphasized. A notable exception is \cite{Donley:dg} which proposes methods to improve privacy by minimizing the leakage between the zones based on adding noise to loudspeaker signals. Other related works rely solely on linear time-invariant filtering and amplitude control \cite{Elliott:2012dv,Cai:2014ih}. The problem we study is different since we do not require silence away from the focusing spots, only unintelligibility. This difference opens up new algorithmic possibilities.

\begin{figure}[t]
\centering
\includegraphics[width=.8\linewidth]{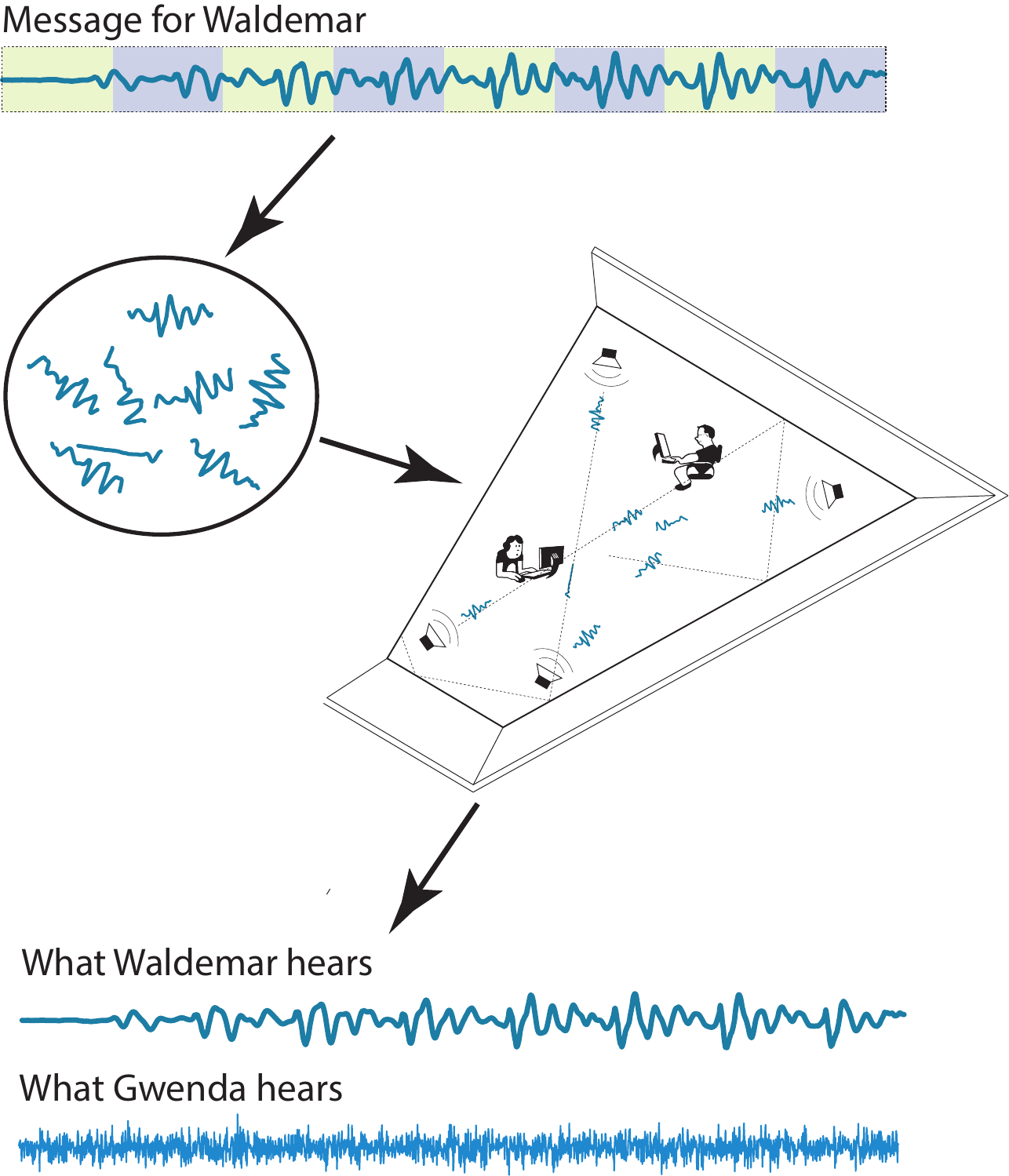}
\caption{Illustration of the proposed method. The audio message is split into short segments which are played out of different loudspeakers after being carefully filtered. The filtering ensures that all the segments neatly align for Waldemar, but Gwenda only hears noise.}
\label{fig:illustration}
\end{figure}

We now perform a thought experiment with reference to Fig. \ref{fig:illustration}: imagine that we divide the message waveform intended for \hbox{Waldemar} into short bursts of sound and emit different bursts from different loudspeakers. Can we somehow delay and filter those short bursts of sound so that at the intended listening point (in this case point A), with the help of echoes, the segments neatly align to form the desired message, while arriving in a disordered, haphazard manner away from A?

In this paper we show that this is indeed possible. The described thought experiment leads to a new method for private audio communication with potential to extend to other modalities such as radio. Interestingly, we show that performance can be improved if instead of emitting short segments of the intended message, we emit filtered Gaussian noise bursts.

In typical soundfield reproduction applications, rendering in reverberant rooms requires additional provisions to deal with the reverberation, with various techniques being employed to handle the echoes \cite{Jin_2015,Betlehem_2005}. In our case, reverberation is precisely what makes the method work---we could not do it without echoes.

Unintelligibility of the messages outside the focusing spots is achieved not by careful filter design, but rather by the choice of loudspeaker driving signals. There is no explicit optimization associated with it.  On the one hand, this makes our method less flexible in terms of controlling the extent and the shape of the sweet spots; on the other hand, the design procedure is very simple: we only need to know the room impulse responses at the points where we want to deliver the messages, not anywhere else.

Another important boon is that unlike many traditional sound focusing methods, such as the time reversal method, \cite{Yon_2003, Chang_2009} and soundfield reproduction approaches, \cite{Betlehem_2015,Berkhout_1993, Ward_2001, Coleman_2014, Wu_2009} our method requires a small number of loudspeakers---we achieve good results with six.

We demonstrate empirically that the proposed method works in real rooms under various model uncertainties. We present simulation results and real experimental results, and evaluate them in terms of the short-time objective intelligibility (STOI) metric. A more persuasive test is to simply listen to the obtained experimental recordings which are available online together with the code to reproduce the results.\footnote{\url{https://swing-research.github.io/sonicdot/}} 

\begin{figure}
\includegraphics[width=.9\linewidth]{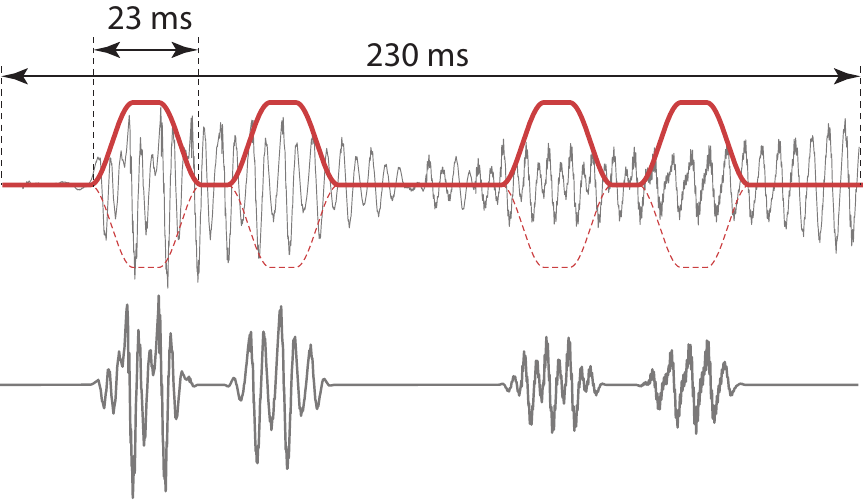}
\caption{Top: original signal $x_k$ multiplied with a mask $w_{k \ell}$ (in red). Bottom: the resulting chopped signal $\tilde{x}_{k\ell}[n].$}
\label{masking}
\end{figure}

\section{Formulation}
\label{sec:formulation}

We explain our formulation with reference to Figure \ref{fig:illustration}. We want to transmit private messages to $K$ users using $L$ loudspeakers placed around the room. This is achieved by randomly splitting the messages intended for different users into chunks, which can be modeled by multiplicative masks (see Fig. \ref{masking}). Denote the message intended for user $k$ by $x_k[n]$. For every $k \in \set{1, \ldots, K}$ we produce $L$ masks $\set{w_{k\ell}}_{\ell=1}^L$ and assign the masked signals $\tilde{x}_{k\ell}[n] = x_k[n] w_{k \ell}[n]$ to the $\ell$th loudspeaker after adequate LSI filtering. We refer to $\tilde{x}_{k\ell}$ as the \emph{design signal}.

\subsection{Mask Design} 
\label{sub:mask_design}

Multiplicative masks are designed to segment every user message into $L$ submessages assigned to each of the $L$ loudspeakers. If the chopped segments are too short or the window used to divide the signal into segments is discontinuous, the recombined message will contain unpleasant audible artifacts. This is partly due to the non-ideal electroacoustical response of the loudspeakers.

A better idea is to segment the signals by smooth, overlapping windows that rise and fall over $T$ samples, and flatten out over $D$ samples. An example of such a smooth mask is illustrated in Fig. \ref{masking}. The transition has a cosine profile so that 
\[
    w[n]
    = 
    \begin{cases}
        \cos^2 \left[ \frac{\pi}{2} \left( 1 - \frac{n}{T-1} \right)  \right], & 0 \leq n < T \\
        1, & T \leq n < T + D \\
        \cos^2 \left[ \frac{\pi}{2} \left( 1 - \frac{n - (T+D)}{T-1} \right)  \right], & T + D \leq n < 2T+D.
    \end{cases}
\]
which is a variation on the Tukey window.

We generate $L$ such smooth masks ensuring that they sum to a constant, 
\begin{equation}
    \label{eq:sum_to_one}
    \sum_{\ell=1}^L w_{k\ell}[n] = 1, \quad \forall k \in \set{1, \ldots, K}, n \in \set{0, \ldots, N-1}.
\end{equation}
The logic behind \eqref{eq:sum_to_one} is that in the anechoic case, simply reproducing adequately delayed and amplified signals would achieve the desired effect since it ensures that
\(
    \sum_{\ell = 1}^L \tilde{x}_{k \ell}[n] = x_k[n].
\)

\begin{figure*}[t!]
\centering
\includegraphics[width=.9\linewidth]{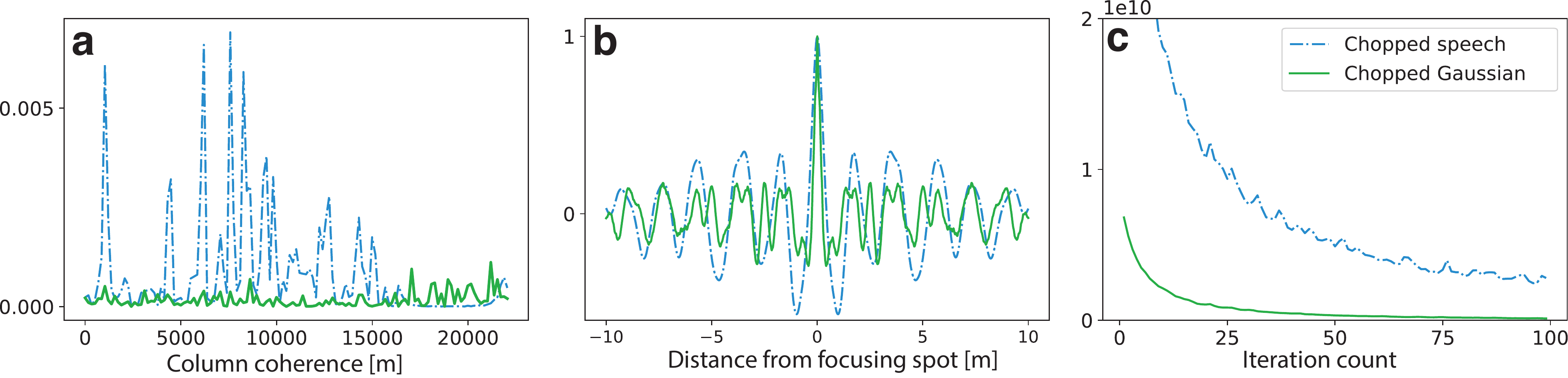}
\caption{Analysis metrics: (a) column coherence; (b) autocorrelation of the loudspeaker driving signal $s_\ell[n]$; (c) residual decay.}
\label{fig:metrics}
\end{figure*}

\subsection{Designing the Spot Filters} 
\label{sub:filter_design}

Instead of directly reproducing $\tilde{x}_{k\ell}[n]$, we first filter it by $g_{k \ell}[n]$. The role of $g_{k\ell}[n]$ is to adjust the phases of emitted chunks of sound so that after echoing about the room they align at the sweet spots. Designing these filters is the main computational step in our method.

Each loudspeaker emits filtered masked signals for each of the $K$ users. Denoting the signal emitted by the $\ell$th loudspeaker by $s_\ell[n]$, we write
\[
    s_\ell[n] 
    = \sum_{k = 1}^K \tilde{x}_{k\ell}[n] \conv g_{k \ell}[n]
    = \sum_{k = 1}^K \big\{x_k[n] w_{k \ell}[n]\big\} \conv g[n].
\]
Denote the room impulse response between the $\ell$th loudspeaker and the $k$th user by $h_{k \ell}[n]$. Then the $k$th user hears the following combined signal:
\begin{align*}
    y_k[n] 
    = \sum_{\ell=1}^L s_\ell[n] \conv h_{k \ell}[n] 
    = \sum_{\ell=1}^L \sum_{k' = 1}^K \tilde{x}_{k'\ell}[n] \conv g_{k'\ell}[n] \conv h_{k \ell}[n].
\end{align*}
The design goal is to make $y_k$ as similar as possible to (a delayed version of) $x_k$. 

We consider finite-length signals of $N$ samples and define the following vectors:
\begin{equation*}
\begin{aligned}
    &\vx_k       & \bydef & \quad \big[ &&x_k[0]      ,&&  x_k[1], &&\ldots, &&x_k[N-1] &\big]^\T, \\
    &\vw_{k\ell} & \bydef & \quad \big[ &&w_{k\ell}[0],&&  w_{k\ell}[1], &&\ldots, &&w_{k\ell}[N-1] &\big]^\T, \\
    &\vg_{k\ell} & \bydef & \quad \big[ &&g_{k\ell}[0],&&  g_{k\ell}[1], &&\ldots, &&g_{k\ell}[M-1] &\big]^\T, \\
    &\vh_{k\ell} & \bydef & \quad \big[ &&h_{k\ell}[0],&&  h_{k\ell}[1], &&\ldots, &&h_{k\ell}[P-1] &\big]^\T.
\end{aligned}
\end{equation*}
We further let $\tilde{\mX}_{k\ell}$ be the $(N+M-1) \times M$ Toeplitz matrix that corresponds to a linear convolution of $\vx_k$ and a signal of length $M$,
\[
    \setlength\arraycolsep{3.2pt}
    \tilde{\mX}_{k\ell} =
    \begin{pmatrix}
        \tilde{x}_{k\ell}[0] & 0 & 0 & 0 & \cdots & 0  \\
        \tilde{x}_{k\ell}[1] & \tilde{x}_{k\ell}[0] & 0 & 0 & \cdots & 0 \\
        \tilde{x}_{k\ell}[2] & \tilde{x}_{k\ell}[1] & \tilde{x}_{k\ell}[0] & 0 & \cdots & 0 \\
        \vdots & \ddots & \ddots & \vdots & \cdots & 0 \\
        0 & 0 & 0 & \cdots & \tilde{x}_{k\ell}[N-1] & \tilde{x}_{k\ell}[N-2] \\
        0 & 0 & 0 & \cdots & 0 & \tilde{x}_{k\ell}[N-1]
    \end{pmatrix},
\]
or
\(
    \tilde{\mX}_{k\ell} = \toeplitz\big( \tilde{\vx}_{k\ell} \big) = \toeplitz\big(\diag(\vw_{k\ell}) {\vx}_k  \big)
\)
for short. The signal driving the $\ell$th loudspeaker can then be written as
\[
    \vs_\ell = \sum_{k=1}^K \tilde{\mX}_{k\ell} \vg_{k \ell} = \tilde{\mX}_\ell \vg_\ell,
\]
where $\tilde{\mX}_\ell=\big[\, \tilde{\mX}_{1\ell}, \tilde{\mX}_{2\ell}, \ \ldots, \tilde{\mX}_{K\ell}  \,\big]$,   $\vg_\ell = \big[\, \vg_{1\ell}^\T, \ \vg_{2\ell}^\T, \ \ldots, \vg_{K\ell}^\T  \,\big]^\T$. Finally, the $k$th user receives
\[
    \vy_k = \sum_{\ell=1}^L \mH_{k\ell} \vs_\ell = \mH_k \tilde{\mX} \vg,
\]
with $\mH_{k\ell} = \toeplitz(\vh_{k\ell}$), $\tilde{\mX} = \mathsf{blockdiag} \big( \tilde{\mX}_1, \ldots, \tilde{\mX}_L \big)$, and $\vg = \big[\, \vg_1^\T, \ \vg_2^\T, \ \ldots, \vg_L^\T  \, \big]^\T$. Collecting all $K$ users in a single matrix--vector equation, we get
\[
    \vy = \mH \tilde{\mX} \vg,
\]
with $\vy = \big[\, \vy_{1}^\T, \ \vy_{2}^\T, \ \ldots, \vy_{K}^\T  \,\big]^\T$, $\mH = \big[\, \mH_{1}^\T, \ \mH_{2}^\T, \ \ldots, \mH_{K}^\T  \,\big]^\T$.

The task is to find the long filter vector $\vg \in \R^{(MLK)\times 1}$ which we estimate by linear least squares,
\begin{equation}
    \label{eq:loss}
    \hat{\vg} = \argmin_{\vg} \ \norm{\vec{\xi} - \mH \tilde{\mX} \vg}_2^2,
\end{equation}
where $\vec{\xi}$ is the delayed version of $\vx$. The solution is in principle given as $\hat{\vg} = (\mH \tilde{\mX})^\dag \vec{\xi}$, where $(\cdot)^\dag$ denotes the Moore-Penrose pseudoinverse. However, the involved matrices are far too large for the naive computation of the pseudoinverse. Instead, we use the conjugate gradient method. Since both $\mH \tilde{\mX}$ and the adjoint $\tilde{\mX}^\T \mH^\T$ consist of multiplications by convolution matrices, the conjugate gradient method can be efficiently implemented using fast Fourier transforms.

\section{Analysis of the Method}

In this section we empirically evaluate several metrics that affect performance in terms of intelligibility in the focusing spots, (lack of) intelligibility outside the focusing spots, and cross-talk between the spots. 
The purpose is to understand why using chopped noise as the design signal considerably outperforms using chopped speech. For actual intelligibility measures the reader may want to fast forward to Section \ref{sec:experimental_results}.

\subsection{Coherence of the System Matrix}

If the computed filters $g_{k\ell}[n]$ are poorly conditioned (i.e., they have both very large and very small coefficients), any model mismatch such as minute changes in room impulse responses will result in large errors in the signals received by the users. We can expect to get unsatisfactory filter responses when the matrix $\mH\tilde{\mX}$ is poorly conditioned.

As a proxy to conditioning which has a useful signal processing meaning, we use frequency-dependent coherence between randomly chosen pairs of columns in $\mH \tilde{\mX}$. For two signals $z[n]$, $w[n]$, coherence is defined as
\[
    \gamma_{zw}(f) = \frac{\abs{C_{zw}(f)}^2}{A_{zz}(f) A_{ww}(f)},
\]
with $C_{zw}$ being the Fourier transform of the crosscorrelation of $z$ and $w$, and $A_{zz}$, $A_{ww}$ Fourier transforms of their autocorrelations.

$\mH \tilde{\mX}$ has $L$ blocks of columns, each corresponding to one loudspeaker. Columns in the same block are correlated as they are influenced by the impulse responses and driving signals; it is desirable that the columns in different blocks be incoherent. In Fig. \ref{fig:metrics}a we plot coherence between columns from \emph{different} blocks. It is clear that using chopped speech as the design signal $\tilde{x}_{k\ell}$ gives a coherent $\mH\tilde{\mX}$ at many frequencies, while using chopped noise gives low coherence.

On the other hand, we observe empirically that as soon as $\mH \tilde{\mX}$ has at least as many columns as rows, its row rank is full and the system has at least one solution. This happens when
\[
    \underbrace{K(P + N + M - 2)}_{\text{length of}~\vy} \leq \underbrace{K M L}_{\text{length of}~\vg} \Longrightarrow M(L - 1) \geq P + N - 2,
\]
that is, as soon as the filters are long enough and we have sufficiently many loudspeakers. This has a nice interpretation: in principle, we can obtain any target signal by multichannel filtering of chopped noise. With chopped speech the matrix $\mH \tilde{\mX}$ is near-singular and the result is brittle.

\subsection{Decay of the Autocorrelation}

Another interesting metric is the decay of the autocorrelation of loudspeaker driving signals. It is a proxy to how fast the sound will decorrelate and become unintelligible as we move away from the focusing points. Fig. \ref{fig:metrics}b, shows that using chopped noise as the design signal $\tilde{x}_{k\ell}$ yields the fastest decay of the autocorrelation. To understand why, note that the autocorrelation $a_{s_\ell s_\ell}[n]$ of the emitted signal $s_\ell[n]$ can be written as 
\begin{equation*}
\begin{aligned}
 s_\ell[n] \conv &s_\ell[-n]
    \   = \ \sum_{k, k'} \tilde{x}_{k\ell}[n] \conv g_{k\ell}[n] \conv \tilde{x}_{k'\ell}[-n] \conv g_{k' \ell}[-n] \\
    \ &  = \ \sum_k a_{\tilde{x}_{k\ell} \tilde{x}_{k\ell}}[n] \conv a_{g_{k\ell} g_{k\ell}}[n] 
    + \sum_{k \neq k'}  c_{\tilde{x}_{k\ell} \tilde{x}_{k'\ell}}[n] \conv c_{g_{k\ell} g_{k'\ell}}[n]
    \end{aligned}
\end{equation*}
The crosscorrelation $c_{\tilde{x}_k \tilde{x}_{k'}}[n]$ will depend on the signals used to feed the loudspeakers. In particular, we can expect that if we use  noise, these crosscorrelations will be small thus reducing the overall autocorrelation of the loudspeaker driving signals.

\subsection{Decay of the Residual}

As a consequence of the above, with a fixed number of conjugate gradient iterations we get results of varying quality at the focusing spots. Fig. \ref{fig:metrics}c shows the value of the loss \eqref{eq:loss}. At any given iteration, the approximation by using speech as design signal is much worse than when we use chopped noise. 

\section{Experimental Results} 
\label{sec:experimental_results}

We use the STOI metric \cite{Taal_2010} to quantitatively assess performance of the proposed method. STOI scores range from zero to one with higher scores indicating higher intelligibility. We test two scenarios: forming $\tilde{x}_{k\ell}$ from chopped speech and forming $\tilde{x}_{k\ell}$ from chopped white Gaussian noise.

\begin{figure}[t]
\centering
\includegraphics[width=\linewidth]{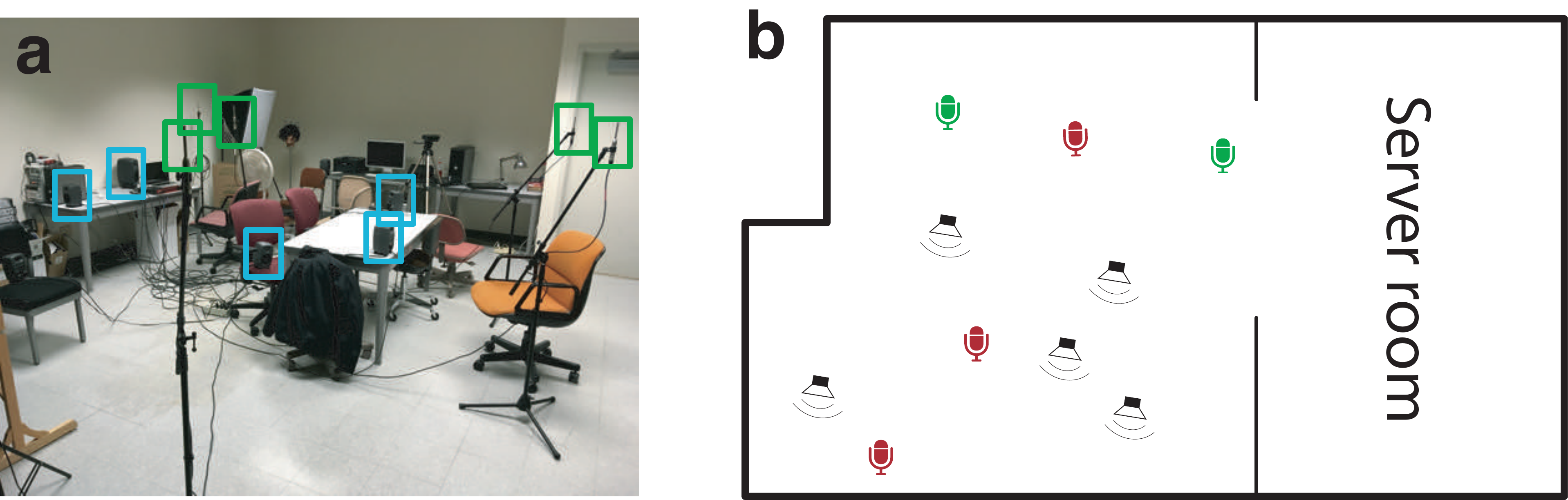}
\caption{Experimental setup. Loudspeakers are placed in an ad hoc manner. In (a), blue rectangles show loudspeaker positions, green rectangles show microphones positions.}
\label{fig:experiment-setup}
\end{figure}

In all experiments we use six loudspeakers placed haphazardly around the room  of size approximately 10 m $\times$ 6 m; see Fig. \ref{fig:experiment-setup}. We report the STOI scores at $K=2$ focusing spots and at three other randomly chosen control locations in the room. The task is to simultaneously deliver two four-second long speech signals to two distinct locations.

\subsection{Numerical Experiments} 
\label{sub:numerical_experiments}

In numerical experiments we use real recorded RIRs but do the convolutions numerically so that unlike in the real experiment there is no model mismatch between measuring the responses and testing the algorithm. The RIRs are measured using the exponential sine sweep technique \cite{Farina_2000} with $L = 6$ loudspeakers---two Genelec 8030B and four Genelec 8010A loudspeakers. We use Audix omnidirectional condenser microphones.

\begin{figure}[htb!]
\includegraphics[width=\linewidth]{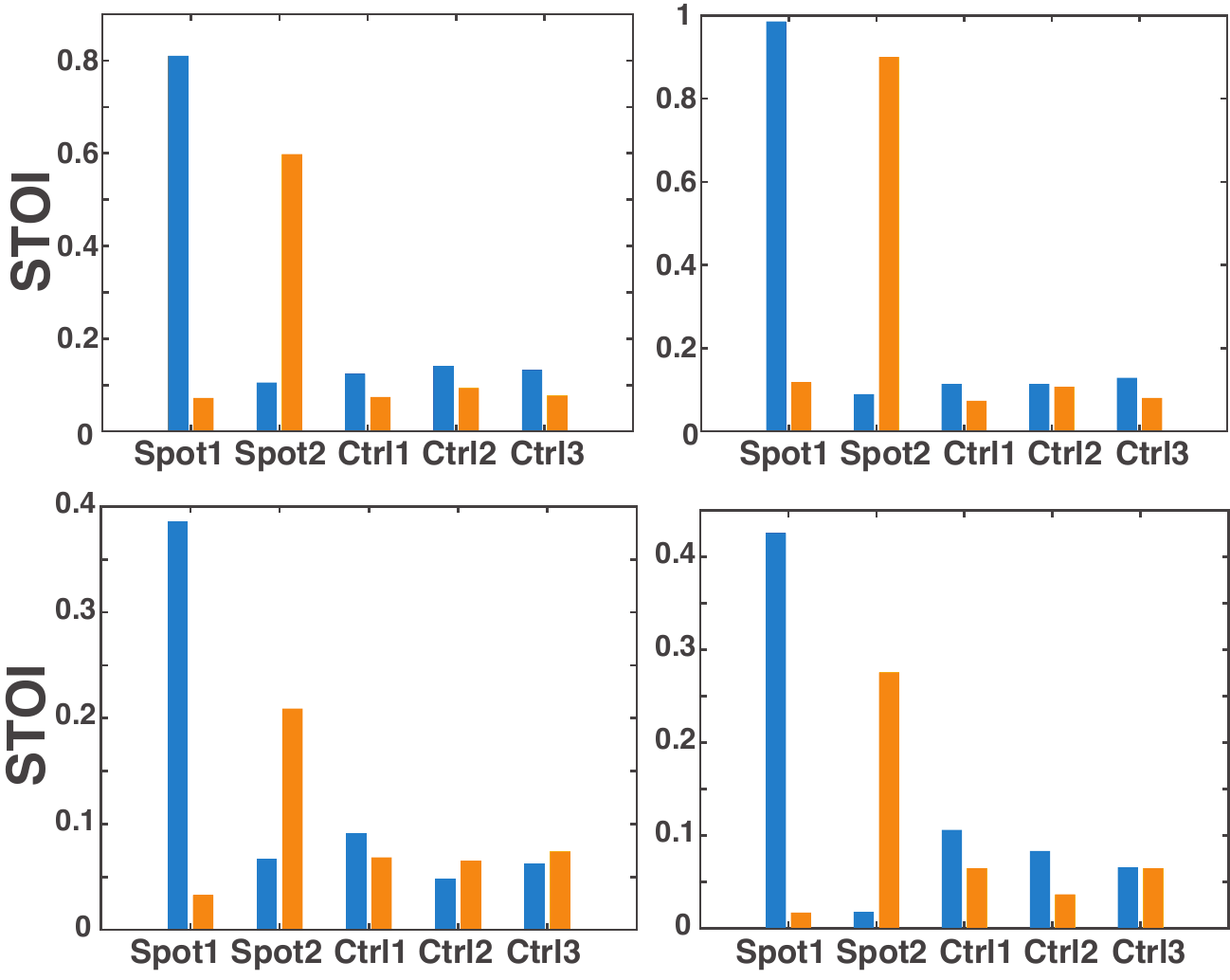}
\caption{Intelligibility measure in numerical experiments (top) and real experiments (bottom), shown at two focusing spots and three control locations. Design signals: left,  speech; right,  noise.}
\label{fig:stoi}
\end{figure}

STOI scores for the five locations are shown in the top row of Fig. \ref{fig:stoi} for the two design signals. We find that for both design signals, significant intelligibility contrast is achieved at the two designated delivery locations and that low intelligibility scores are obtained at the reference points, exactly as desired.

When using noise, the metrics significantly improve. The intelligibility contrast at the two private audio delivery locations is improved, and very low STOI scores are attained at all reference points for both speech signals, suggesting that neither message is intelligible. Informal listening experiments corroborate the quantitative observations. As evident from online sound samples, chopped noise is superior to chopped speech in meeting the design goals.

\subsection{Experiments in a Real Room  } 
\label{sub:experiments_in_a_real_room_}

We conducted real experiments in a laboratory of size approximately 10 m $\times$ 6 m shown in Fig. \ref{fig:experiment-setup}, the same room where the RIRs for the simulation were collected. The number of loudspeakers was $L = 6$, the number of sweet spots $K = 2$, and we used three control points. We note that between the time the room responses were collected and the time the method was tested, people moved about the room, and small objects and chairs were moved about. Other parameters such as the temperature were also likely altered due to air conditioning (they were not controlled). In Fig. \ref{fig:stoi}, bottom, the STOI scores are shown for the two design signals. Again, using chopped noise performs better than using chopped speech, though the scores are overall lower due to the model mismatch between the measuring time and testing time.

\section{Conclusion} 
\label{sec:conclusion}

We presented a new method for private audio messaging. The gist of the method is in emitting filtered chunks of sound from a relatively small number of loudspeakers which then get recombined by echoes in just the right way at just the right points. The proof-of-concept experiments show that the method performs well, rendering highly-intelligible speech with inaudible cross-talk at the focusing points, and ``junk'' at other points.

One drawback of the current method is that a new set of filters must be designed for every combination of input signals. Interesting future work is to study how to streamline filter design when messages change, ideally in real time. Our analysis is at the moment on a phenomenological level---a more fundamental understanding of the method is necessary. Ongoing work includes a quantification of the effect of the number of loudspeakers, quality of loudspeakers, and room shape and clutter. Another interesting question is whether we can algorithmically control the shape and the extend of the focusing spot. Finally, it seems clear that the proposed method can be applied to any wave modality as well as a number of other problems where focusing and incoherence are required simultaneously.

\section{Acknowledgment}

We would like to thank Mihailo Kolund\v{z}ija who first told us about the thought experiment and Robin Scheibler for early discussions about this work.

\bibliographystyle{IEEEbib}
\bibliography{ref_IWAENC_2018}

\end{document}